\documentclass[11pt]{article}
\usepackage{jheppub}
\usepackage{amssymb}
\usepackage{amsmath}
\usepackage{amsthm}
\usepackage{graphicx}
\usepackage{float}
\title{CP violating phase from charged-lepton mixing}
\author[a]{J. Alberto Acosta,}
\author[b,c]{Alfredo Aranda,}
\author[b]{Julio Virrueta}

\affiliation[a] {PRISMA Cluster of Excellence, Institut f\"ur Physik, Johannes Gutenberg-Universit\"at Mainz, D - 55099 Mainz, Germany}
\affiliation[b] {Facultad de Ciencias - CUICBAS, Universidad de Colima, C.P. 28045, Colima, M\'exico}
\affiliation[c] {Dual CP Institute of High Energy Physics, C.P. 28045, Colima, M\'exico}
 
\emailAdd{acosta@uni-mainz.de}
\emailAdd{fefo@ucol.mx}
\emailAdd{jmoreno9@ucol.mx}

\abstract{
A model independent analysis of the leptonic Dirac CP-violating phase ($\delta$) is presented. The analysis uses the
experimentally determined values of the mixing angles in the lepton mixing matrix in order to explore the allowed values for $\delta$ and possible general forms for the charged lepton mixing matrix. This is done under two general assumptions: 1) that the mixing matrix in the neutrino sector is the so-called tribimaximal matrix and hence the non zero value for $\theta_{13}$ arises due to the mixing matrix in the charged lepton sector and 2) the charged lepton mixing matrix is parametrized in terms of three angles and one phase. It is found that any value of $\delta$ is still consistent with the data and that, considering the assumptions above,  regardless of the value for $\delta$, the $1-3$ mixing angle in the charged lepton sector is small but non zero and the $2-3$ mixing angle can take values in only two possible small ranges around $0$ and $\pi/2$ respectively.}


\notoc

\begin{document}

\maketitle
%%%%%%%%%%%%%%%%%%%%%RESUMEN%%%%%%%%%%%%%%%%%%%%%%%%%%%%%%%%%%%%%%%%%%%%%%%%%%%%%%%%%
%%%%%%%%%%%%%%%%%%*******************%%%%%%%%%%%%%%%%%%%%%%%%%%%%%%%%%%%%%%%%%%%%%%%%
%%%%%%%%%%%%%%%%%%%%%INTRODUCCION%%%%%%%%%%%%%%%%%%%%%%%%%%%%%%%%%%%%%%%%%%%%%%%%%%%%
%%%%%%%%%%%%%%%%%%*******************%%%%%%%%%%%%%%%%%%%%%%%%%%%%%%%%%%%%%%%%%%%%%%%%
\section{Introduction}
The measurement of a {\it sizeable} $\theta_{13}$ mixing angle in the lepton sector~\cite{An:2012eh,Ahn:2012nd,Abe:2011sj} has opened up the very interesting possibility of exploring the Dirac CP violating phase (hence CP violation) present in the Pontecorvo-Maki-Nakagawa-Sakata (PMNS) mixing matrix~\cite{Fritzsch:1995dj}. On a first stance this non-zero value excludes the so-called tribimaximal (TBM) pattern for the PMNS matrix~\cite{tbm}, where $\theta_{13} = 0$, that led to an extensive search for flavor symmetries and mechanisms which reproduced it. On the other hand, the TBM structure is still perfectly consistent as a mixing matrix of the neutrino sector if the charged lepton sector is no longer diagonal, as it is commonly assumed, and "lifts" the zero in the $1-3$ sector of the PMNS matrix~\cite{Xing:2002sw,pcol}.

What is the structure of the charged lepton mixing matrix $U_{\ell}$? We do not know. Recall that $U_{PMNS} \equiv U_{\ell}^{\dagger}U_{\nu}$, where $U_{\ell}$ and $U_{\nu}$ are the charged lepton and neutrino mixing matrices, respectively. Thus if $U_{\ell} = {\mathbf{1}}$, $U_{PMNS} = U_{\nu} \neq U_{TBM}$. If we insist on having $U_{\nu} = U_{TBM}$, then $U_{\ell}$ must differ from $\mathbf{1}$ in such a way that the following experimentally obtained ranges are obtained:
%current experimental data at $3\sigma$ from the gobal data analysis is \cite{vexp}
\begin{equation}\label{data}
\begin{tabular}{c|cc|ccc}
\hline
 & Best fit value (NH)&  $3\sigma$ range  &  Best fit value (IH) &  $3\sigma$ range \\
\hline
$\sin^{2}\theta_{12}$ & $0.320$ & $0.27-0.37$ &  $0.320$ & $0.27-0.37$ \\
$\sin^{2}\theta_{13}$ &  $0.0246$ & $0.017-0.033$ &  $0.025$ & $0.017-0.033$\\
$\sin^{2}\theta_{23}$ &  $0.514$ & $0.361-0.667$ &  $0.60$ & $0.37-0.67$ \\ \hline
\end{tabular}
\end{equation}
where the values for $\sin^{2}\theta_{12}$ and $\sin^{2}\theta_{13}$ (for both hierarchies) were obtained from~\cite{Tortola:2012te}. For $\sin^{2}\theta_{23}$ we use the  most recent T2K result $\sin^2\theta_{23}=0.514 \pm 0.082$ ($90\%$ C.L.)~\cite{t2k} that is consistent with the Ice Cube Collaboration value~\cite{ice}. In addition to the mixing angles, the recent result on electron neutrino appearance at T2K has provided an exclusion at 90\% C.L. of some ranges for $\delta$, namely ($0.19\pi , 0.8\pi)$ for NH and the two regions at $(-\pi , -0.97\pi)$ and $(-0.04\pi, \pi)$ for IH (they use the convention $-\pi \leq \delta \leq \pi$)~\cite{Abe:2013hdq}. Note that for the IH case these results imply that $\sin\delta < 0$.

% There are several studies related to the issue of finding the matrices $U_{\ell}$ that might work. In fact this was studied in different contexts even before the measurement of $\theta_{13}$. 

The purpose of this paper is to discuss, in as general terms as possible, what can be said about the only parameter in the PMNS matrix that has not been measured and that we know is there, regardless of the fermionic nature of neutrinos, namely, the Dirac CP-violating phase $\delta$.  Another goal is to use the available data in order to determine whether or not there are particularly identifiable classes and/or textures for the charged lepton mixing matrix $U_{\ell}$. Some of this has been analysed by some previous works under different perspectives. Antusch and King~\cite{Antusch:2005kw} used the idea of quark-lepton complementarity to explore specific relations between the mixing angles in both sectors in order to make predictions for $\delta$ in the lepton sector. Recently, a very interesting proposal was put forward by D. Marzocca and collaborators~\cite{pet} where specific parametrizations were introduced for the  neutrino and charged lepton sectors in such a way that it is possible, under certain circumstances, to relate angles in both sectors  and thus have a predictive scenario for $\delta$. In a previous work~\cite{pcol} we found that given a {\it Particle Data Group} (PDG)~\cite{Beringer:1900zz} type parametrization for the charged lepton mixing matrix in terms of charged lepton mixing angles $\theta_{ij}^{\ell}$  (with the Dirac CP-violating phase $\delta$ set to zero), it was possible to obtain the PMNS matrix with small $\theta^{\ell}_{12}$ and $\theta^{\ell}_{13}$ while large values of $\theta^{\ell}_{23}$ were required (see~\cite{Frampton:2004ud} for a similar situation with bimaximal neutrino mixing).

\section{Discussion and results}
Motivated by those ideas and our previous results, we explore the scenario where the neutrino mixing matrix is given by the TBM matrix and the charged lepton mixing matrix is parametrized {\it \`a la} PDG, namely
\begin{equation}
 U_{TBM}=\left(\begin{array}{ccc}
 \sqrt{\frac{2}{3}}&\frac{1}{\sqrt{3}}&0\\
 -\frac{1}{\sqrt{6}}& \frac{1}{\sqrt{3}}&-\frac{1}{\sqrt{2}}\\
-\frac{1}{\sqrt{6}}&\frac{1}{\sqrt{3}}&\frac{1}{\sqrt{2}}
 \end{array}\right)
\end{equation}
and
\begin{equation} \label{Ul}
 U_{\ell}=\left(\begin{array}{ccc}%
 c^{\ell}_{12}c^{\ell}_{13} & s^{\ell}_{12}c^{\ell}_{23} & s^{\ell}_{13}e^{-i\delta^{\ell}} \\
 -s^{\ell}_{12}c^{\ell}_{23}-c^{\ell}_{12}s^{\ell}_{23}s^{\ell}_{13}e^{i\delta^{\ell}} & c^{\ell}_{12}c^{\ell}_{23}-s^{\ell}_{12}s^{\ell}_{23}s^{\ell}_{13}e^{i\delta^{\ell}} & s^{\ell}_{23}c^{\ell}_{13} \\
 s^{\ell}_{12}s^{\ell}_{23}-c^{\ell}_{12}c^{\ell}_{23}s^{\ell}_{13}e^{i\delta^{\ell}} & -c^{\ell}_{12}s^{\ell}_{23}-s^{\ell}_{12}c^{\ell}_{23}s^{\ell}_{13}e^{i\delta^{\ell}} & c^{\ell}_{23}c^{\ell}_{13}
 \end{array}\right)
\end{equation}
where $c^{\ell}_{ij}$ and $s^{\ell}_{ij}$ stand for  $\cos\theta^{\ell}_{ij}$ and $\sin\theta^{\ell}_{ij}$ respectively and $\delta^{\ell}$ denotes a phase (note that $\delta^{\ell}=0$ imply $\delta=0$). We do not include the possible Majorana phases in this discussion. Note that this parametrization is the same commonly used for the $U_{PMNS}$ matrix with $\theta^{\ell}_{ij}$ and $\delta^{\ell}$ replaced by $\theta_{ij}$ and $\delta$ respectively. Using the definition $U_{PMNS} \equiv U^{\dagger}_{\ell}U_{TBM}$ one then obtains
\begin{eqnarray}\label{rel1}
\sin^{2}\theta_{13}&=&\frac{\alpha^{2}_{1}+\alpha^{2}_{2}}{2}+\alpha_{1}\alpha_{2}\cos\delta^{\ell} \ , \\\label{rel2}
\sin^{2}\theta_{23}&=&\frac{\alpha^{2}_{2}+\alpha^{2}_{3}-2\alpha_{2}\alpha_{3}\cos\delta^{\ell}}{2(1-\sin^{2}\theta_{13})} \ , \\\label{rel3}
\sin^{2}\theta_{12}&=&\frac{\alpha^{2}_{4}+\alpha^{2}_{5}-2\alpha_{4}\alpha_{5}\cos\delta^{\ell}}{3(1-\sin^{2}\theta_{13})} \ ,
\end{eqnarray}
where
\begin{eqnarray}\label{alphas}
\alpha_{1}&=&\sin\theta^{\ell}_{12}(\sin\theta^{\ell}_{23}+\cos\theta^{\ell}_{23})\ , \\
\alpha_{2}&=&\cos\theta^{\ell}_{12}\sin\theta^{\ell}_{13}(\sin\theta^{\ell}_{23}-\cos\theta^{\ell}_{23})\ , \\
\alpha_{3}&=&\cos\theta^{\ell}_{12}(\sin\theta^{\ell}_{23}+\cos\theta^{\ell}_{23})\ , \\
\alpha_{4}&=&\sin\theta^{\ell}_{12}(\sin\theta^{\ell}_{23}-\cos\theta^{\ell}_{23})+\cos\theta^{\ell}_{12}\cos\theta^{\ell}_{13}\ , \\
\alpha_{5}&=&\cos\theta^{\ell}_{12}\sin\theta^{\ell}_{13}(\sin\theta^{\ell}_{23}+\cos\theta^{\ell}_{23})\ .
\end{eqnarray}

\begin{figure}[ht]
%\begin{center}
\includegraphics[width=8cm]{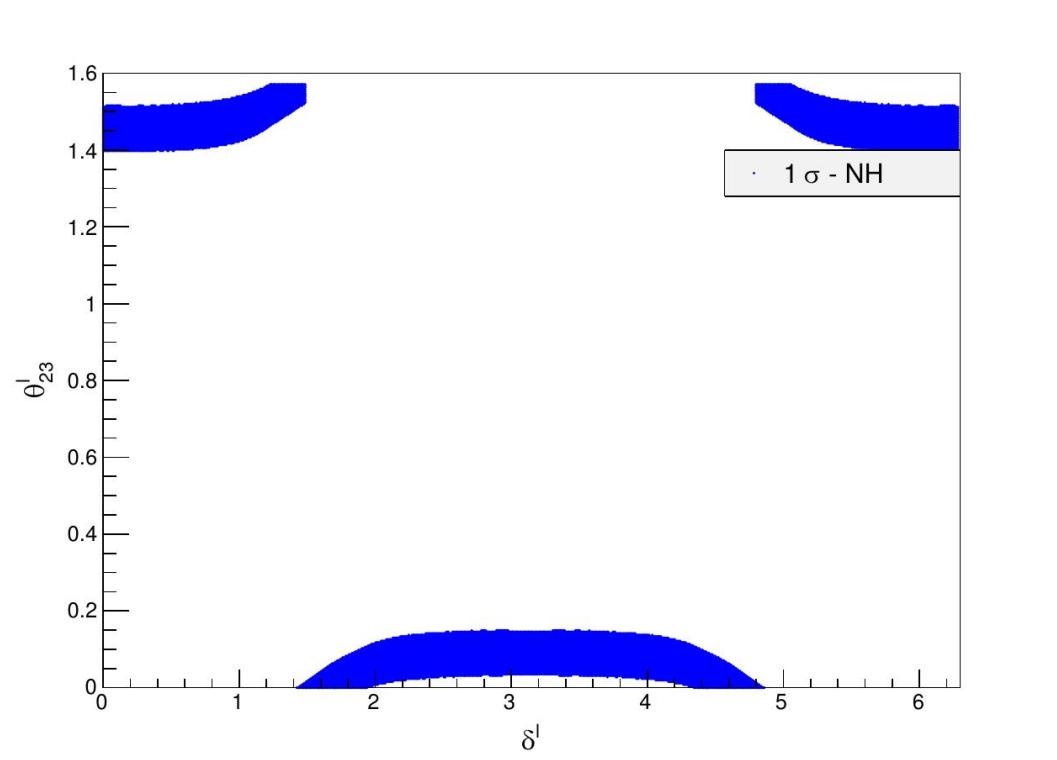} 
\includegraphics[width=8cm]{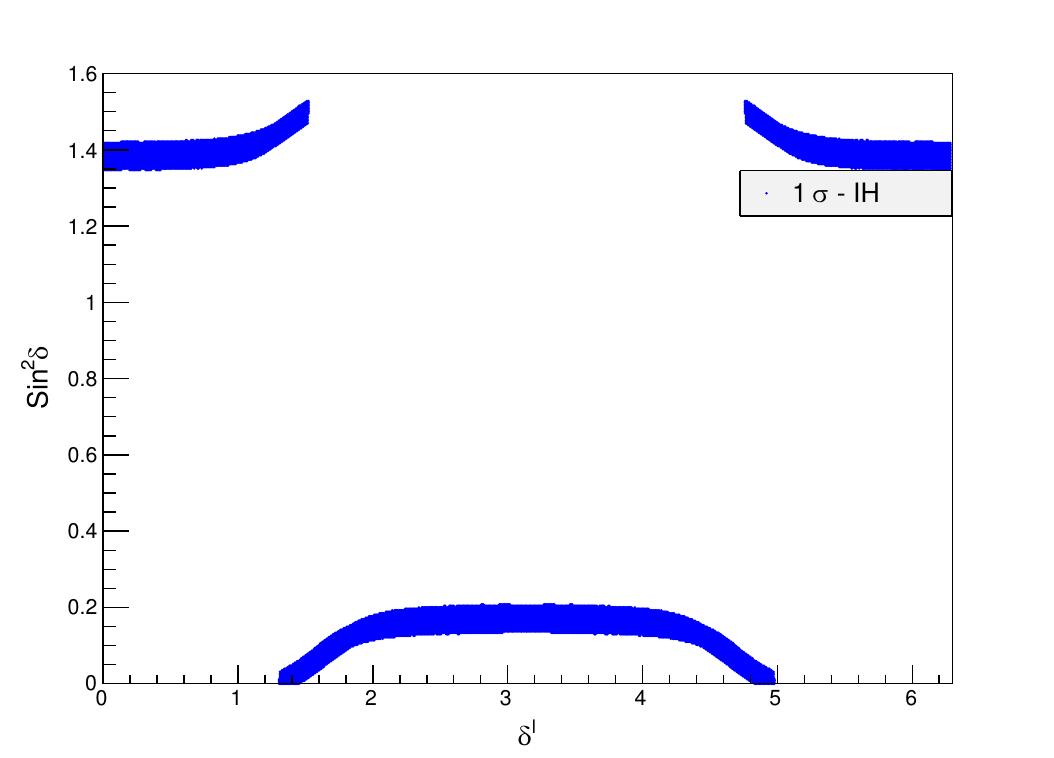} \\
\includegraphics[width=8cm]{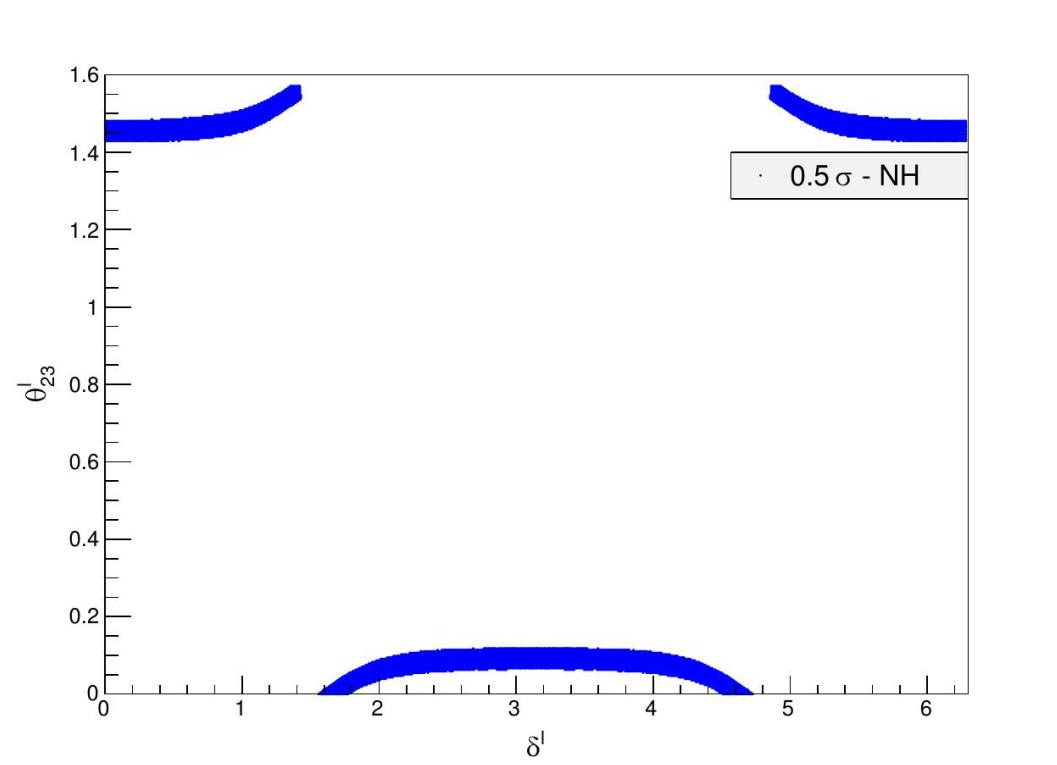} 
\includegraphics[width=8cm]{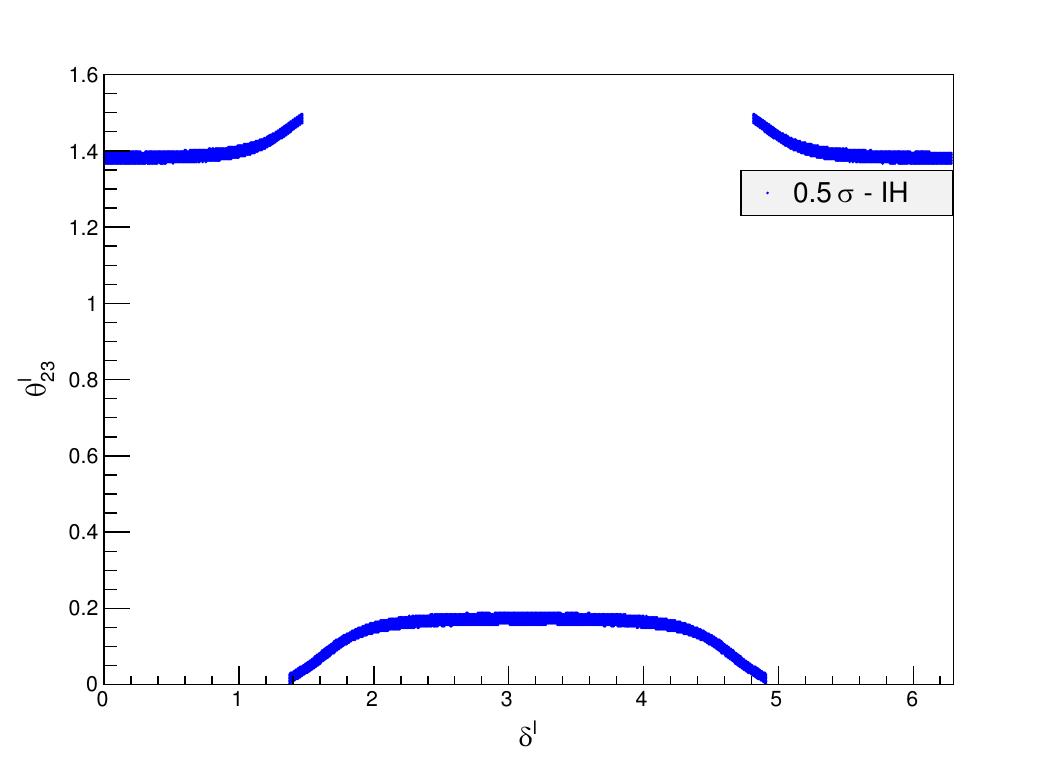} \\
\includegraphics[width=8cm]{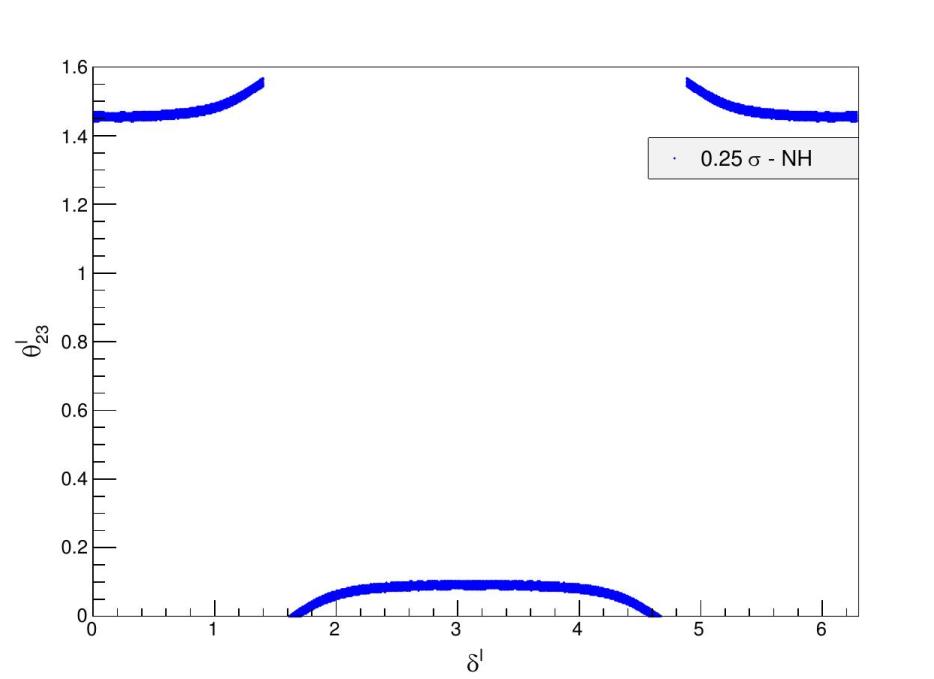} 
\includegraphics[width=8cm]{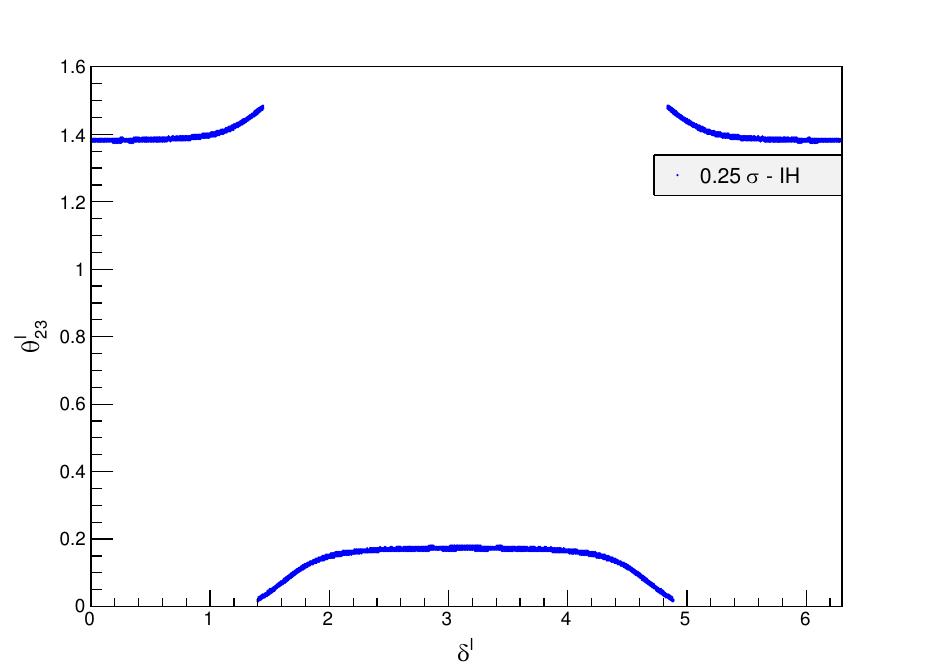} 
%\end{center}
\caption{Evolution of $\theta^{\ell}_{23}$ with respect to $\delta^{\ell}$ 
 for NH in the left column and IH in the right one. We show results taking values consistent with those in Eq.~\eqref{data} at $1\sigma$ (top), $\frac{1}{2}\sigma$ (middle) 
 and very close to the central values at $\frac{1}{4}\sigma$ (bottom). Note that in the NH case, as one gets closer to the central values, there are two small gaps at $\delta^{\ell} \sim \pi/2$ and $\delta^{\ell} \sim 3\pi/2$ where no solutions exist.}
  \label{fig1}
\end{figure}

In the following, all results to be presented correspond to regions in the parameter space of $\theta^{\ell}_{ij}$ and $\delta^{\ell}$ that satisfy the experimental values in Eq.~\eqref{data} at different $\sigma$ levels. An interesting observation found in our previous work~\cite{pcol} was that solutions existed for a range of values corresponding to $\theta^{\ell}_{23} \sim \pi/2$, this for the specific case of  $\delta^{\ell} = 0$, i.e. for cases where no CP violation is present in the lepton sector. This might seem counter intuitive as it corresponds to maximal mixing in the $2-3$ sector of the left-handed charged leptons, something that might not be a priori expected. It turns out however that including the possibility on non-zero values for $\delta^{\ell}$ leads to the situation shown in Figure~\ref{fig1}. There are only two possible value ranges for $\theta^{\ell}_{23}$ distributed in three regions: the two top regions corresponding to values for $\theta^{\ell}_{23} \in (1.4 , \pi/2)$ and the lower one with $\theta^{\ell}_{23} \in (0, 0.15)$ (at the $1-\sigma$ level). The "width" of the three regions correspond to different $\sigma$ values, as described in the caption. It is interesting to note that the regions corresponding to $\delta^{\ell} = 0$ and $\delta^{\ell} = 2\pi$ do correspond to the maximal $\theta^{\ell}_{23} \sim \pi/2$ found in our previous analysis (only for $\delta^{\ell} = 0$) and in fact is valid for the ranges $0 \leq \delta^{\ell} \leq \pi/2$ and $3\pi/2 \leq \delta^{\ell} \leq 2 \pi$, whereas small values of $\theta^{\ell}_{23} \sim 0 - 0.15 (0.2)$ for NH (IH) correspond to the range $\pi/2 \leq \delta^{\ell} \leq 3\pi/2$. This is an interesting result that can have implications in the analysis of  lepton flavor violation in specific models. Note also that the main difference between NH and IH is that, for this last one, we get a slightly higher value for $\theta^{\ell}_{23}$ in the  $\pi/2 \leq \delta^{\ell} \leq 3\pi/2$ range and a bit smaller in the ranges with $\theta^{\ell}_{23} \sim \pi/2$.

\begin{figure}[ht]
%\begin{center}
\includegraphics[width=8cm]{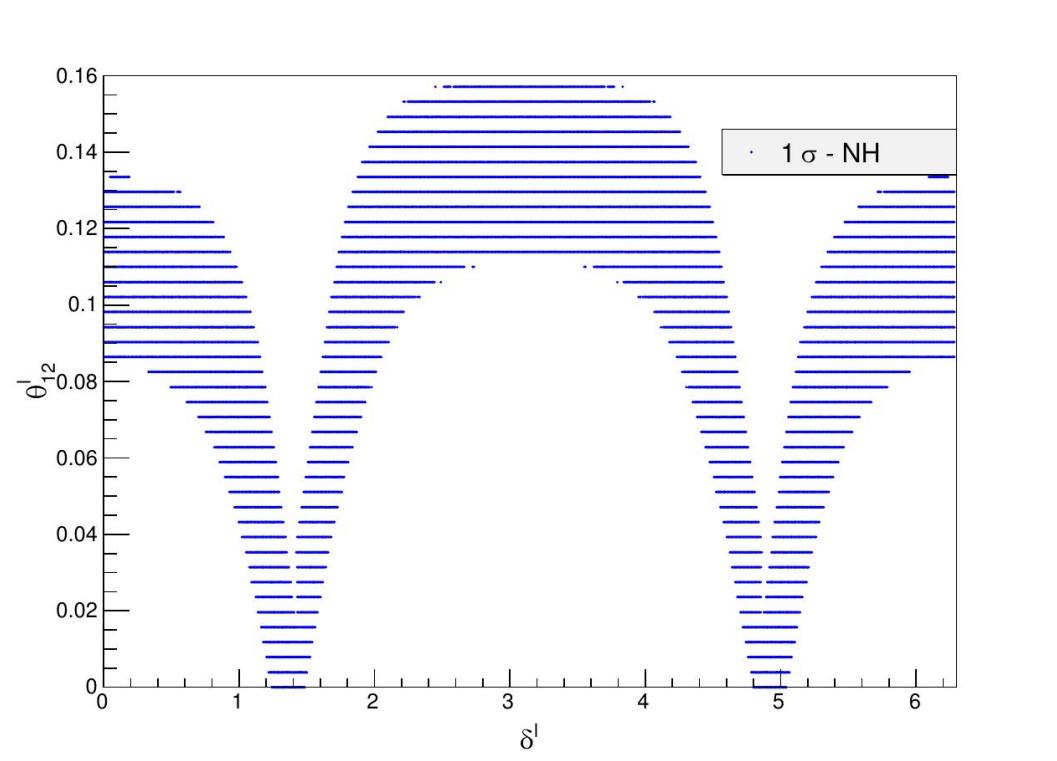} 
\includegraphics[width=8cm]{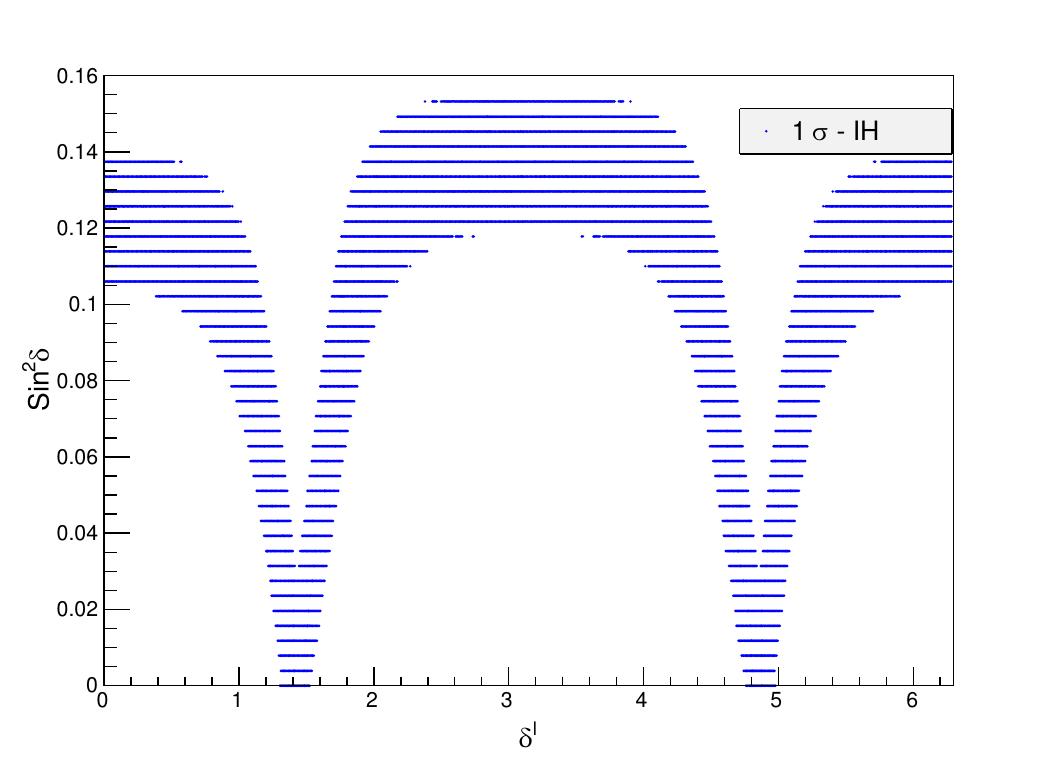} \\
\includegraphics[width=8cm]{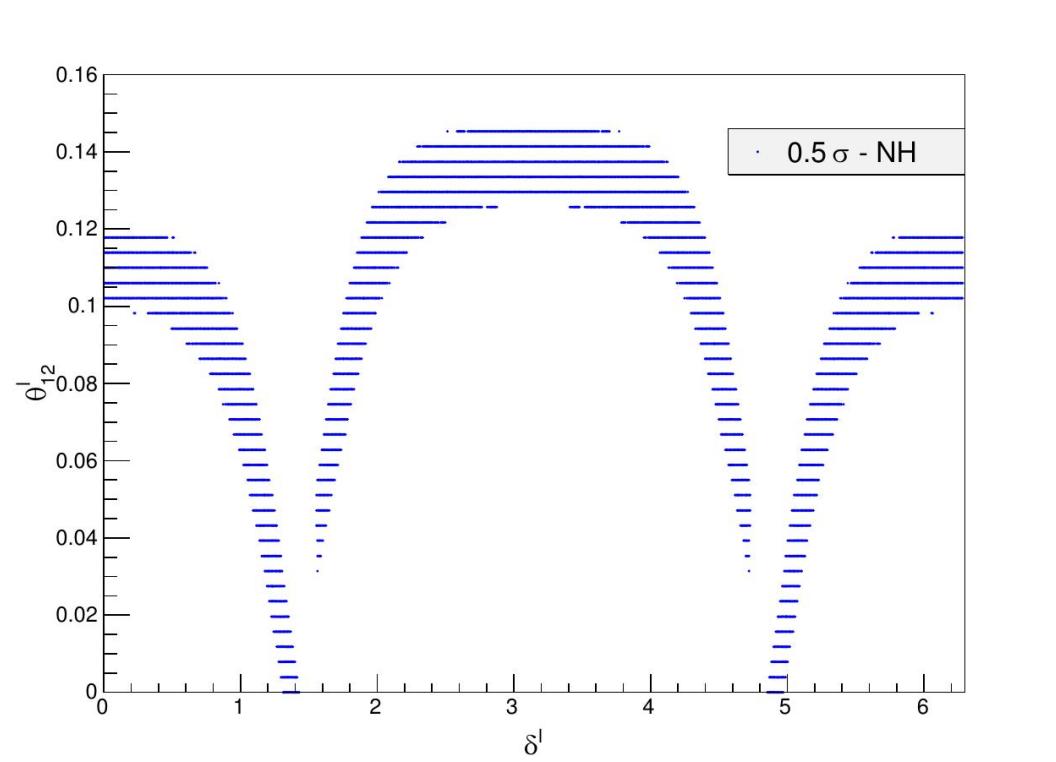} 
\includegraphics[width=8cm]{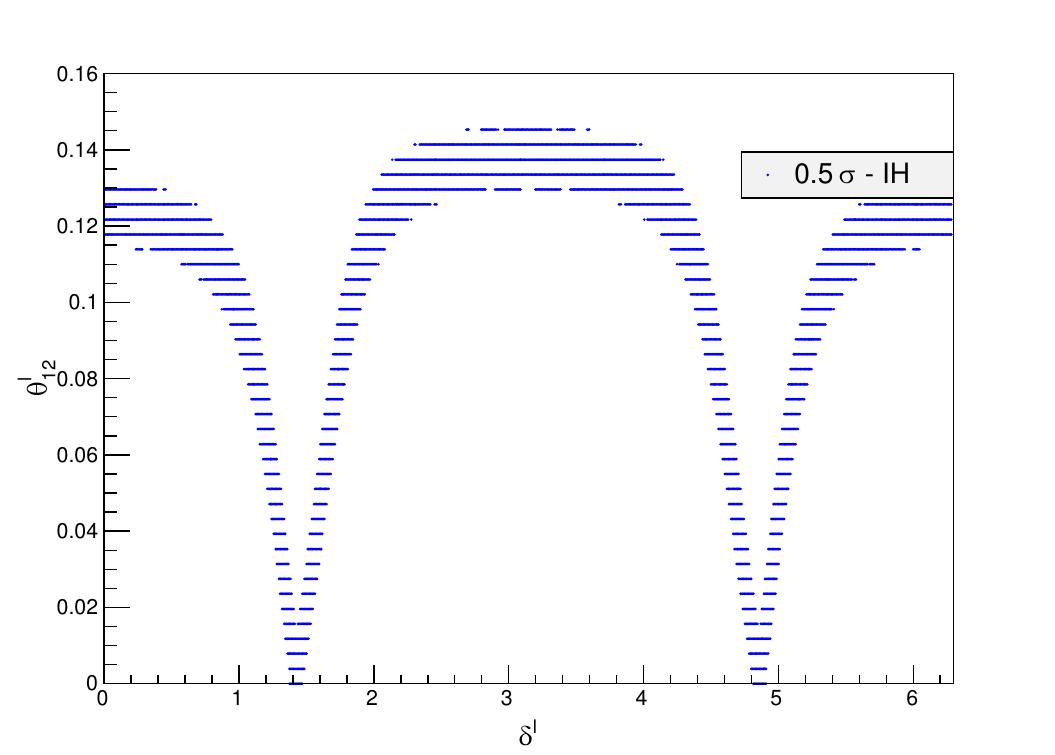} \\
\includegraphics[width=8cm]{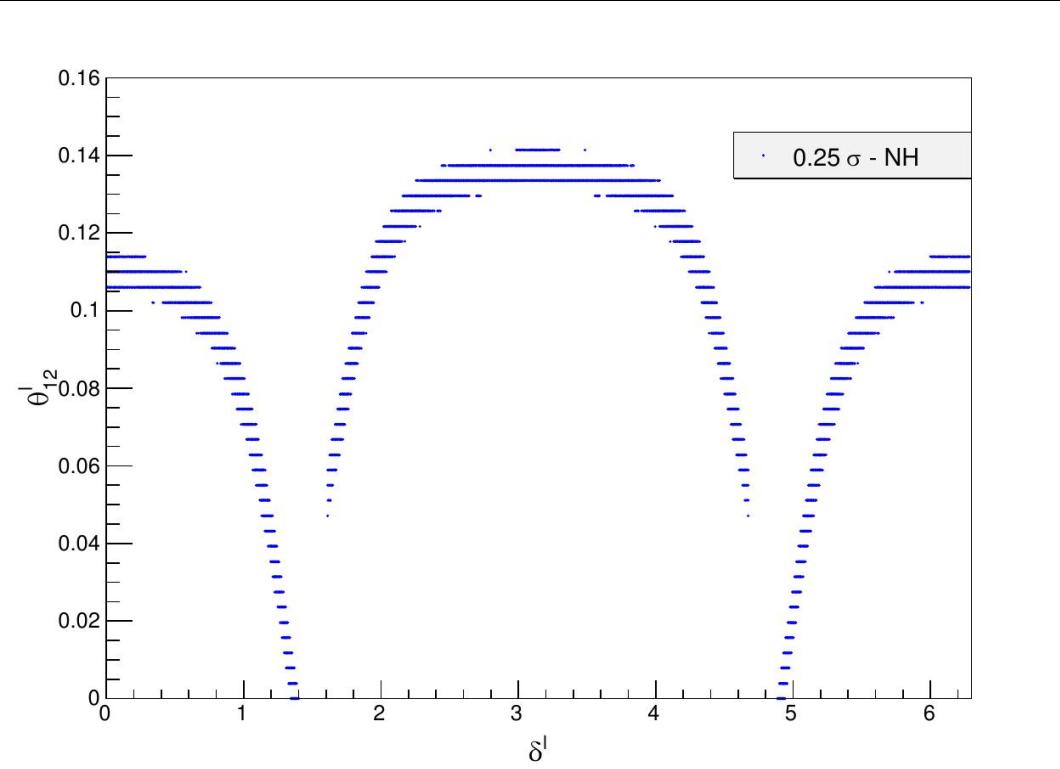} 
\includegraphics[width=8cm]{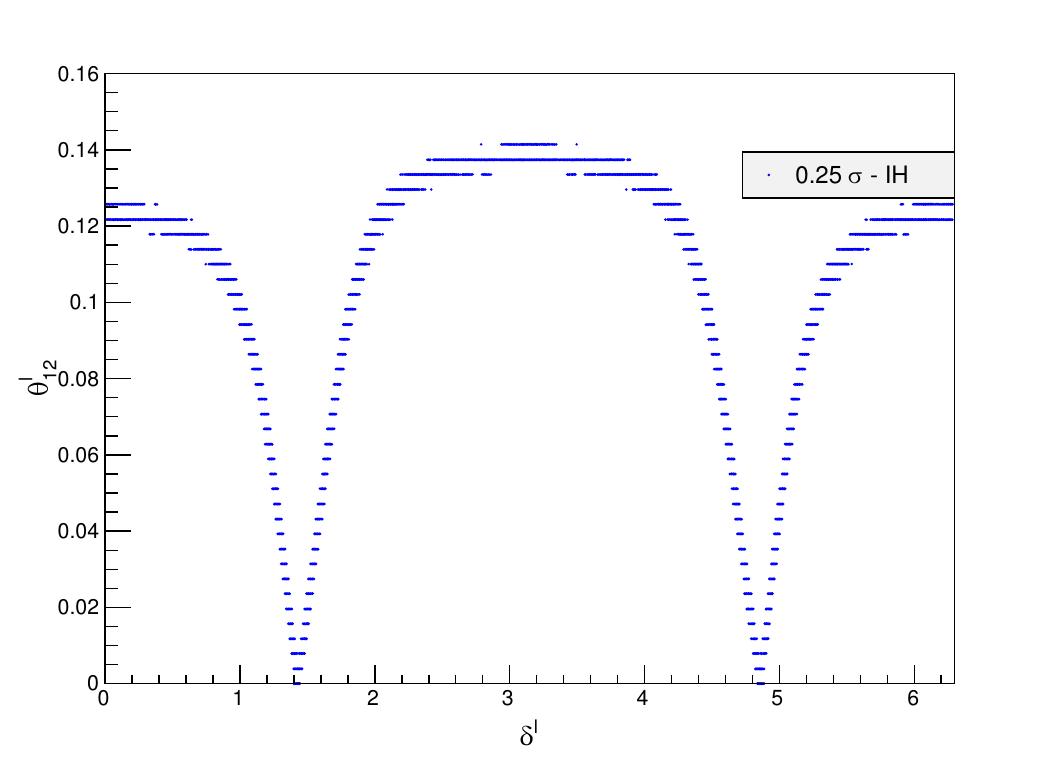} 
 %\end{center}
 \caption{Evolution of $\theta^{\ell}_{12}$ with respect to $\delta^{\ell}$ 
 for NH in the left column and IH in the right one. We show results taking values consistent with those in Eq.~\eqref{data} at $1\sigma$ (top), $\frac{1}{2}\sigma$ (middle) 
 and very close to the central values at $\frac{1}{4}\sigma$ (bottom). Note that as in Figure~\ref{fig1}, for the NH case, as one gets closer to the central values, there are two small gaps at $\delta^{\ell} \sim \pi/2$ and $\delta^{\ell} \sim 3\pi/2$ where no solutions exist.}
  \label{fig2}
\end{figure}

\begin{figure}[ht]
%\begin{center}
\includegraphics[width=8cm]{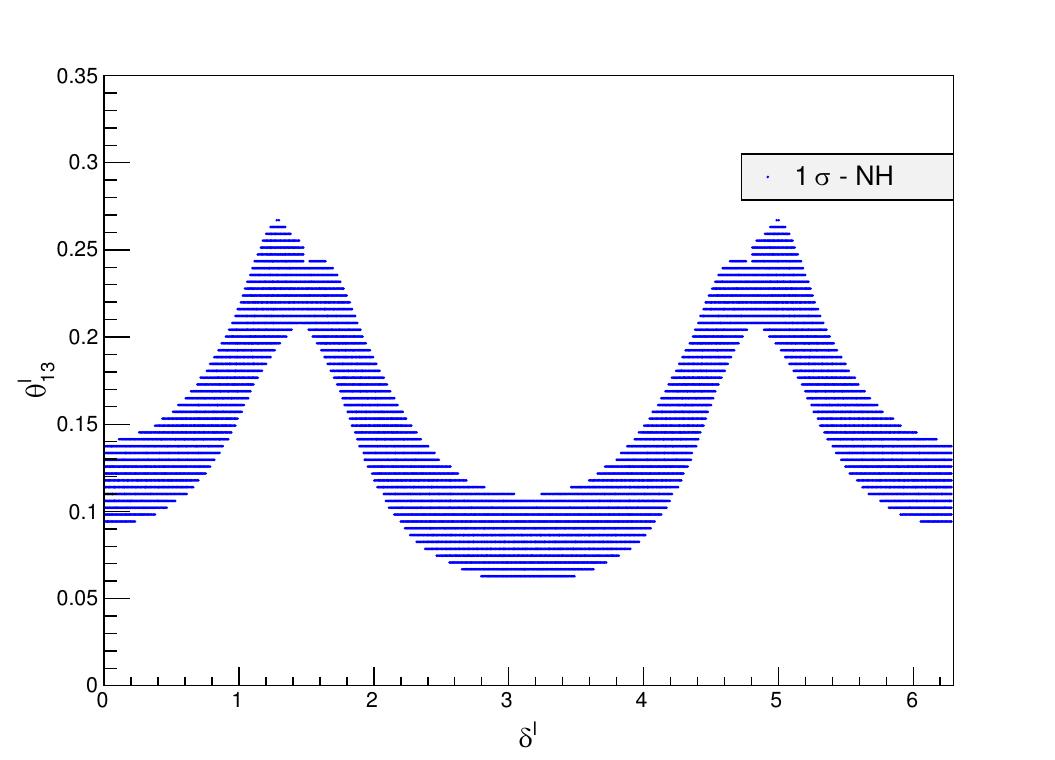} 
\includegraphics[width=8cm]{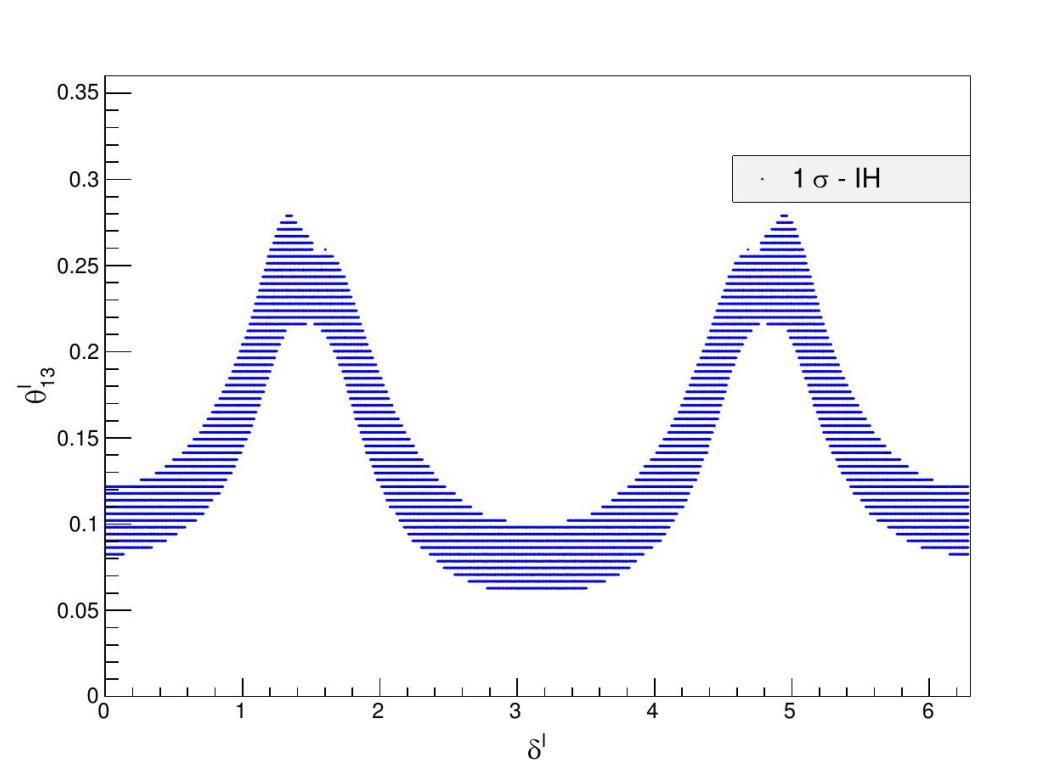} \\
\includegraphics[width=8cm]{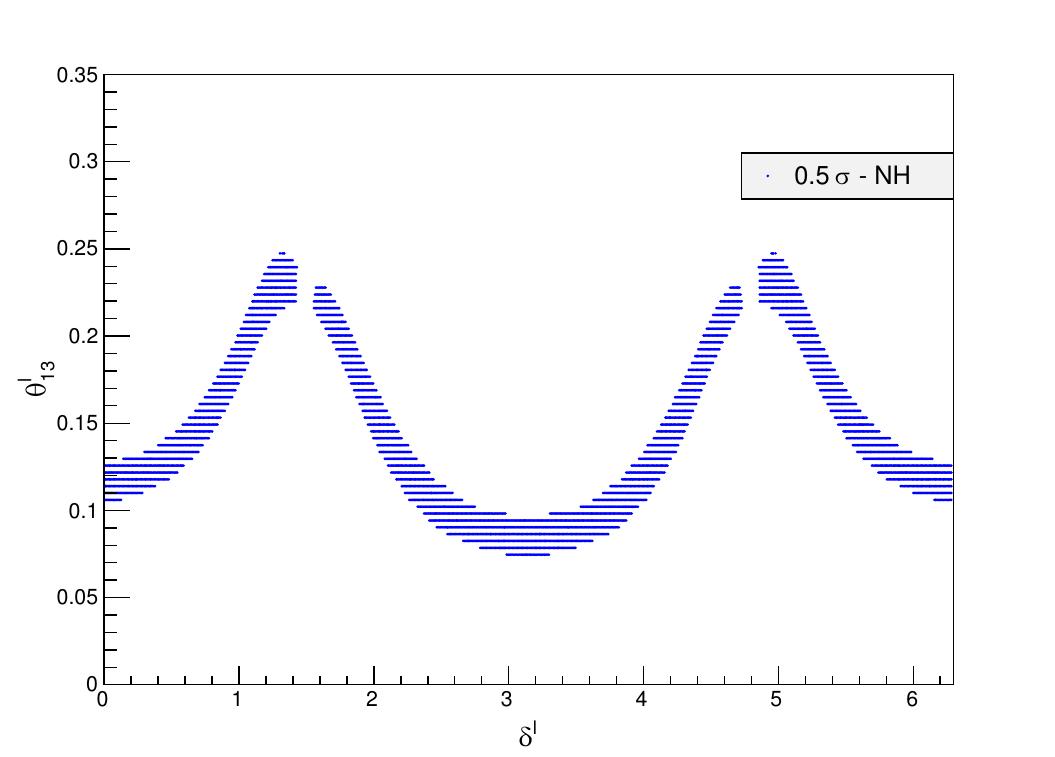} 
\includegraphics[width=8cm]{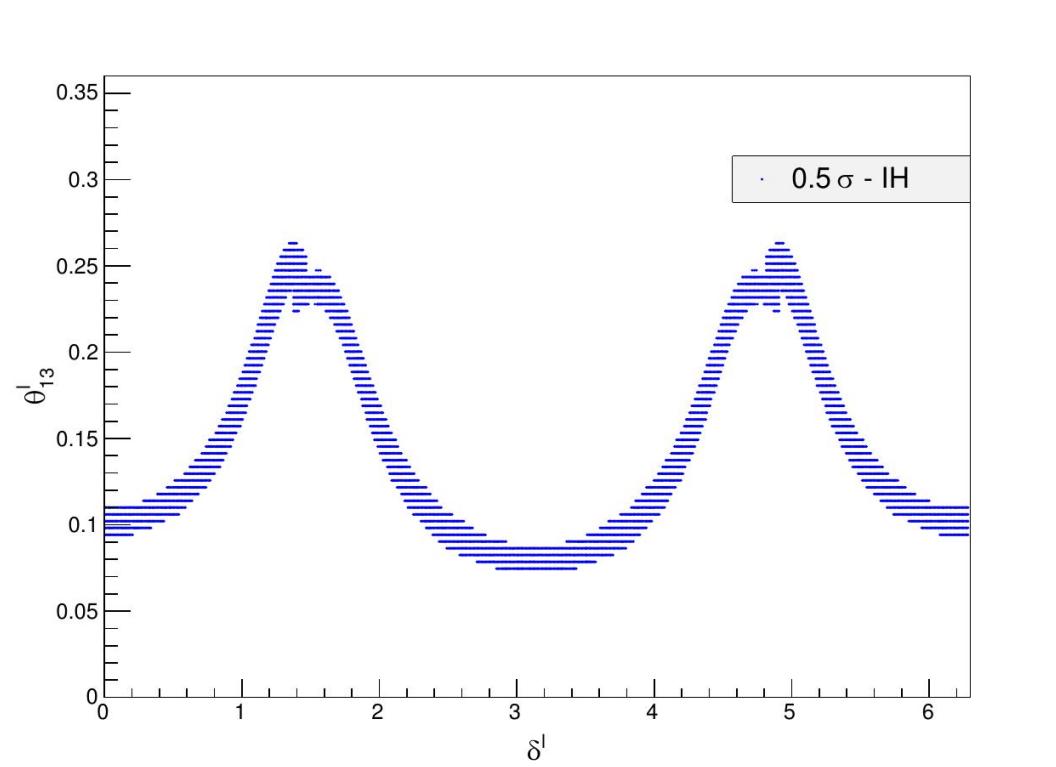} \\
\includegraphics[width=8cm]{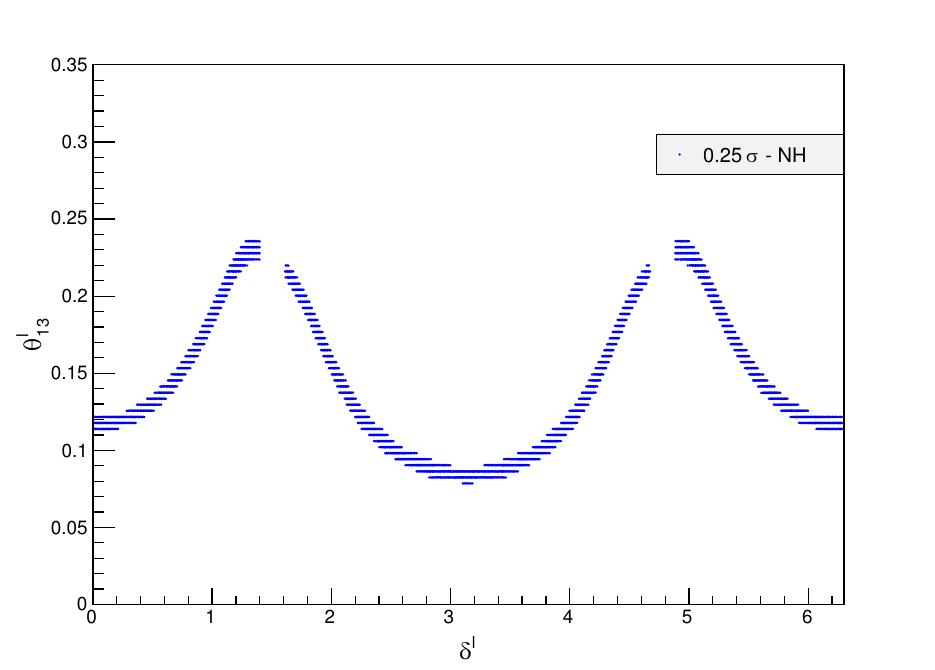} 
\includegraphics[width=8cm]{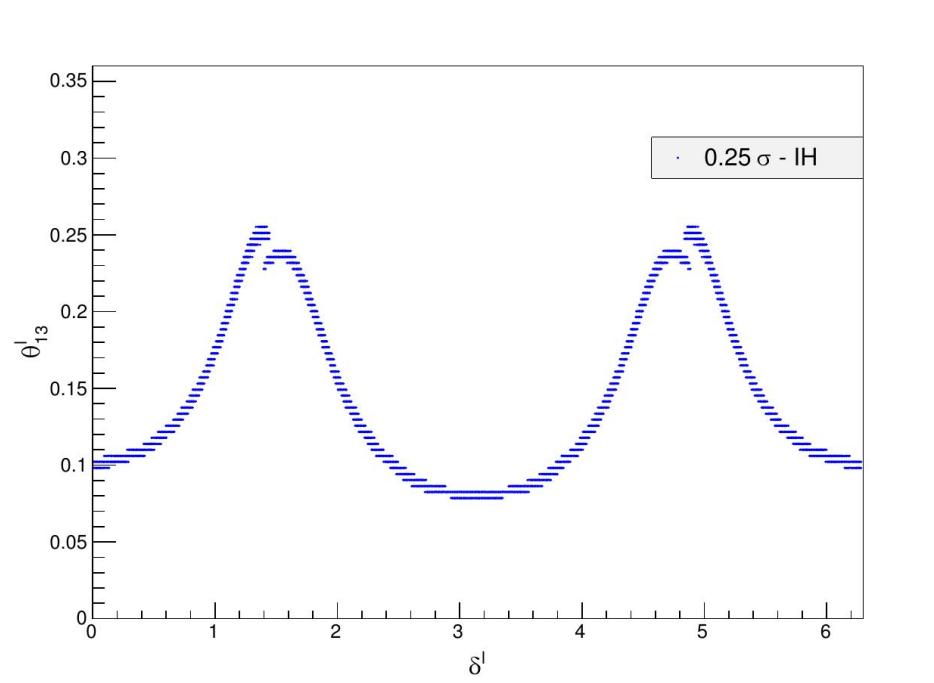} 
 %\end{center}
 \caption{Evolution of $\theta^{\ell}_{13}$ with respect to $\delta^{\ell}$ 
 for NH in the left column and IH in the right one. We show results taking values consistent with those in Eq.~\eqref{data} at $1\sigma$ (top), $\frac{1}{2}\sigma$ (middle) 
 and very close to the central values at $\frac{1}{4}\sigma$ (bottom). Note that $\theta^{\ell}_{13}$  is never zero : this implies that if $\delta^{\ell}\neq 0$ then $\delta \neq 0$. Note that consistent with Figures~\ref{fig1} and ~\ref{fig2}, for the NH case, as one gets closer to the central values, there are two small gaps at $\delta^{\ell} \sim \pi/2$ and $\delta^{\ell} \sim 3\pi/2$ where no solutions exist.}
  \label{fig3}
\end{figure}

The analogous plots for $\theta^{\ell}_{12}$ and $\theta^{\ell}_{13}$ are shown in Figures~\ref{fig2} and~\ref{fig3}, respectively, where it is possible to observe the {\it inverted} correlation between the two angles. Notice that as in the previous case, all values of $\delta^{\ell}$ are allowed, including of course $\delta^{\ell} = 0$. Observe also that the minimum possible value for $\theta^{\ell}_{13}$ in Figure~\ref{fig3} is always greater than zero. This implies that as long as there is a non-zero value for $\delta^{\ell}$, there will be a non-zero CP-violating phase in the PMNS matrix. The allowed values for these angles at the $1-\sigma$ level are $0 \leq \theta_{12}^{\ell} \leq 0.16$  and $0.06 \leq \theta_{13}^{\ell} \leq 0.27$ 

It is clear from these results that it is not possible to obtain a definite prediction - nor any strong hint - on the value of $\delta$ by simply using the currently measured mixing parameters. Yet, we can single out two interesting regions: if one chooses the smallest possible value (at $1\sigma$) for $\theta_{13}^{\ell} \simeq 0.06$, one obtains $\theta_{12}^{\ell} \simeq 0.16$, $\theta_{23}^{\ell} \simeq 0.09$, and $\delta^{\ell} \simeq \pi$. Similarly, taking the smallest possible value for $\theta_{12}^{\ell} \simeq 0$ leads to
$\theta_{13}^{\ell} \simeq 0.28$, $\theta_{23}^{\ell} \in (0 - 0.1)$, and $\delta^{\ell} \simeq \pi/2, 3\pi/2$. These interesting cases lead to charged lepton matrices of the general form
\begin{eqnarray} \label{13min}
U_{\ell}(min(\theta^{\ell}_{13})) &\sim& 
\left(
	\begin{array}{ccc}
	1 - \epsilon^2 & \epsilon & \epsilon^2 e^{-i\delta_{\ell}} \\
	- \epsilon(1+ \epsilon^2 e^{i\delta_{\ell}}) &  1 - \epsilon^{2}(1+ \epsilon^2 e^{i\delta_{\ell}}) & \epsilon^{2} \\
	 \epsilon^{2}(1 - e^{i\delta_{\ell}}) & -\epsilon^{2}(1 + \epsilon e^{i\delta_{\ell}}) & 1 - \epsilon^2 
	\end{array}
\right) \ , \\ \nonumber \\  \label{12min230}
U_{\ell}(min(\theta^{\ell}_{12})) &\sim& 
\left(
	\begin{array}{ccc}
	1 - \epsilon^2 & \epsilon^{4} & \epsilon e^{-i\delta_{\ell}} \\
	- \epsilon^{4}(1+  e^{i\delta_{\ell}}) &  1 - \epsilon^{2} & \epsilon^{4} \\
	 -\epsilon e^{i\delta_{\ell}} & \epsilon^{4}(1 - e^{i\delta_{\ell}}) & 1 - \epsilon^2 
	\end{array}
\right)_{\theta^{\ell}_{23}\sim 0}\ , \\ \nonumber \\  \label{12min2301}
U_{\ell}(min(\theta^{\ell}_{12})) &\sim& 
\left(
	\begin{array}{ccc}
	1 - \epsilon^2 & \epsilon^{4} & \epsilon e^{-i\delta_{\ell}} \\
	- \epsilon^{2}(\epsilon^{2}+  e^{i\delta_{\ell}}) &  1 - \epsilon^{2} & \epsilon \\
	 \epsilon(\epsilon^{3}- e^{i\delta_{\ell}}) & -\epsilon(1 +\epsilon^{3} e^{i\delta_{\ell}}) & 1 - \epsilon^2 
	\end{array}
\right)_{\theta^{\ell}_{23}\sim \epsilon}\ ,
\end{eqnarray}
where $\epsilon ~ O(10^{-1})$.

Note that if one assumes a CKM-like form for the charged lepton mixing matrix~\cite{Plentinger:2005kx,Goswami:2009yy}, i.e. if one assumes (thinking only in terms of relative size and order) $\theta^{\ell}_{12}/\theta^{\ell}_{13} \sim O(10^{2})$,  $\theta^{\ell}_{12}/\theta^{\ell}_{23} \sim O(10^{1})$, then the only possible values for $\delta^{\ell}$ fall in the central region close to $\pi$, depending on the specific values chosen for the angles. This is not particularly relevant aside from the prejudice-driven expectation that the charged lepton mixing matrix could somehow resemble the mixing in the quark sector, as some works have suggested in the past.

Lastly we compute $\sin^{2}\delta$ using the Jarlskog invariant $J_{CP}$ that in standard parametrization is given by
\begin{equation}\label{j}
 J_{CP}=Im\{U^{*}_{e1}U^{*}_{\mu3}U_{e3}U_{\mu1}\}=\frac{1}{8}\sin\delta\sin2\theta_{13}\sin2\theta_{12}\sin2\theta_{23}\cos\theta_{13} \ .
\end{equation}
Every term but $\sin\delta$ in Eq.~(\ref{j}) can be numerically evaluated using Eqs.~\eqref{rel1}, \eqref{rel2}, \eqref{rel3} and the numerical solutions we have computed for all $\delta^{\ell}$. On the other hand, computing
\begin{equation}
  J^{\prime}_{CP}=Im\{U'^{*}_{11}U'^{*}_{23}U'_{13}U'_{21}\}
\end{equation}
for $U'=U^{\dagger}_{\ell}U_{TBM}$ with $U_{\ell}$ given in Eq.~\eqref{Ul}. Again, all the entries in $U'$ can be computed using the solution volume of $\theta^{\ell}_{ij}$ for all $\delta^{\ell}$.  Hence, numerically and without approximations we compute 
\begin{equation}\label{sdc}
\sin^{2}\delta=\frac{( J'_{CP})^{2}}{\sin^{2}\theta_{12}(1-\sin^{2}\theta_{12})\sin^{2}\theta_{13}(1-\sin^{2}\theta_{13})^{2}\sin^{2}\theta_{23}(1-\sin^{2}\theta_{23})} \ .
\end{equation}

\begin{figure}[ht]
%\begin{center}
 \includegraphics[width=8cm]{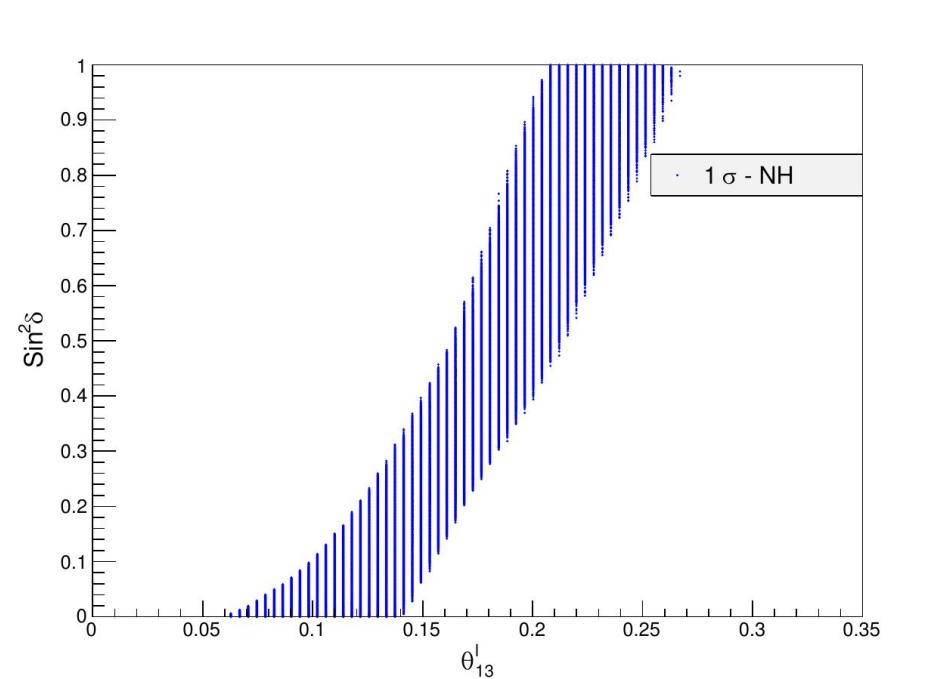}
 \includegraphics[width=8cm]{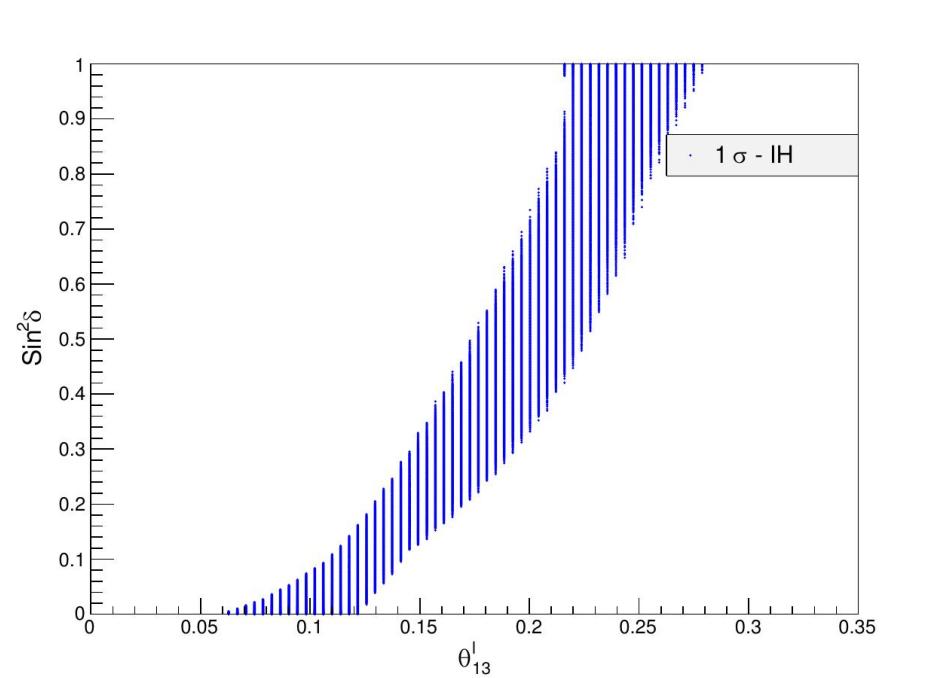} \\
 \includegraphics[width=8cm]{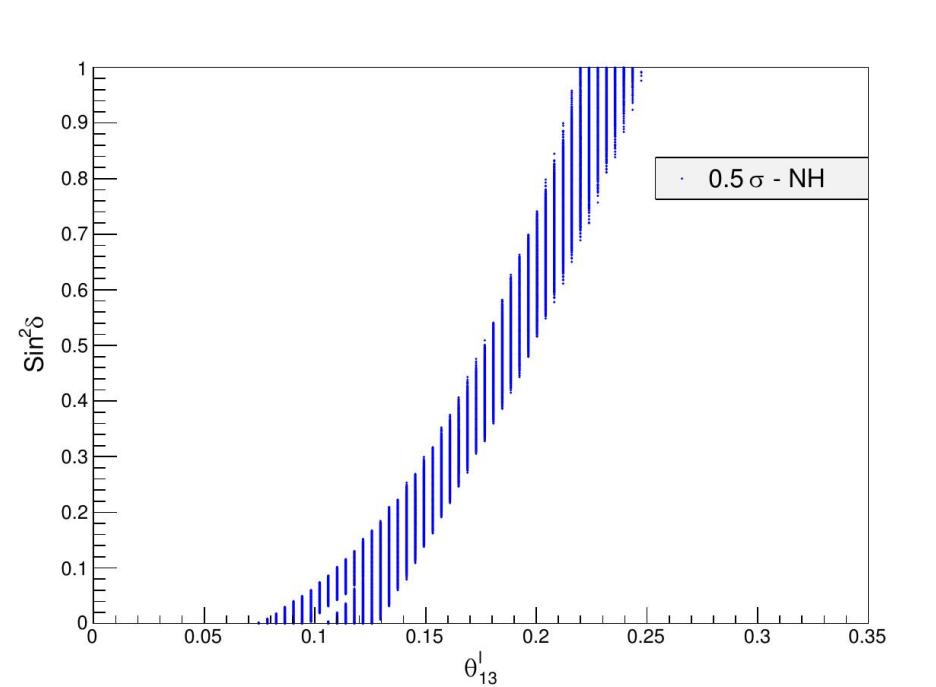}
 \includegraphics[width=8cm]{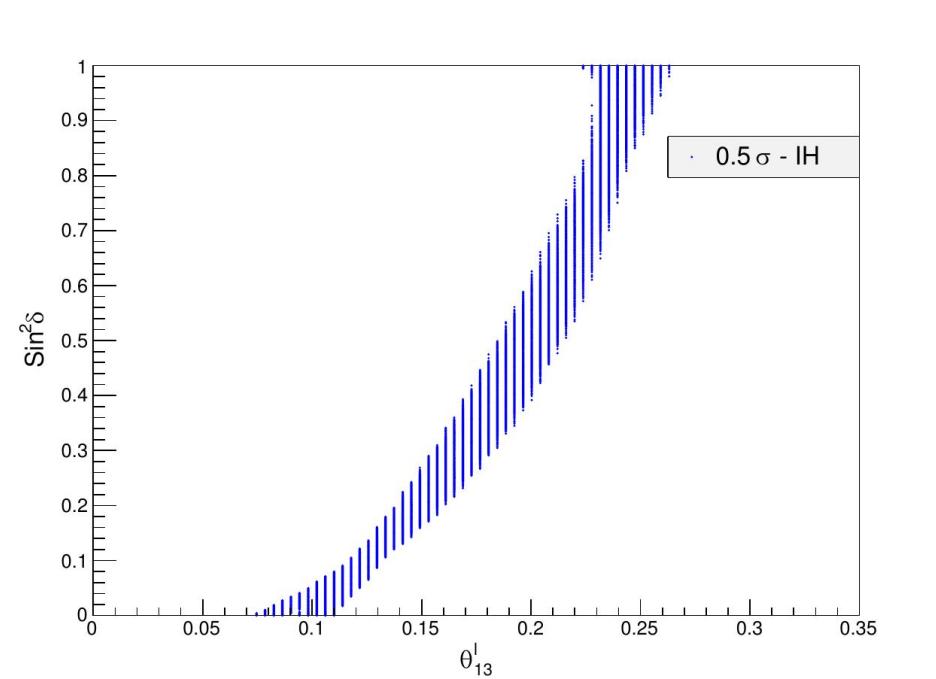} \\
 \includegraphics[width=8cm]{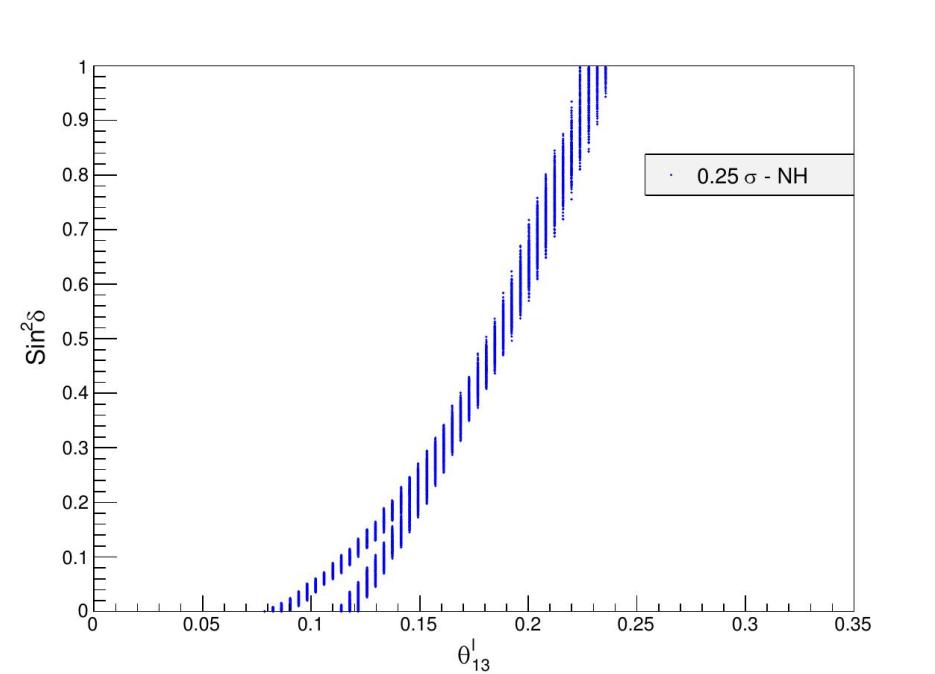}
 \includegraphics[width=8cm]{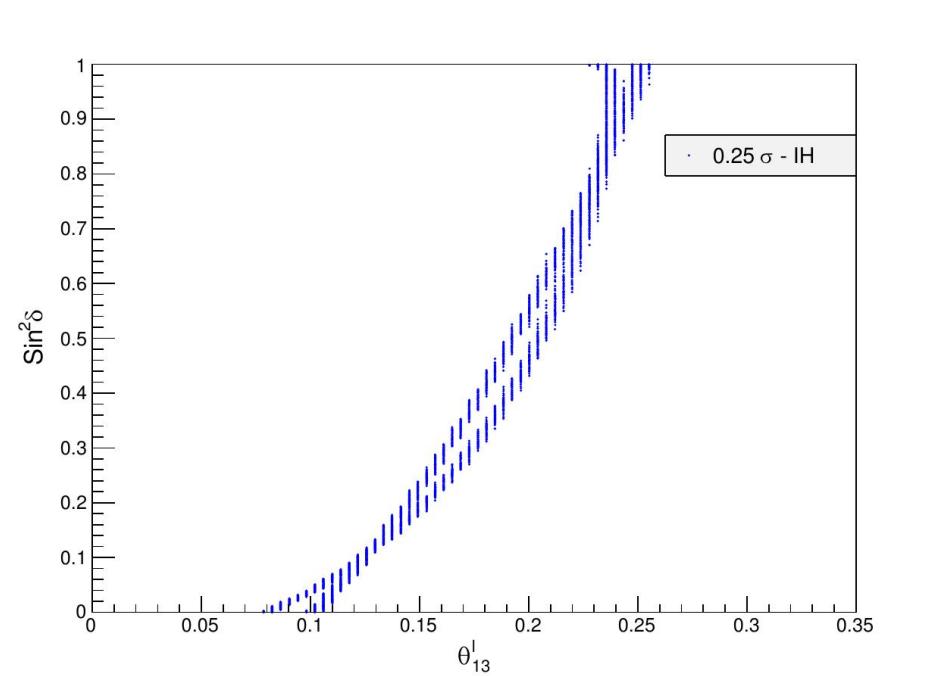}
 %\end{center}
 \caption{Variation of $\sin^{2}\delta$ with respect to $\theta^{\ell}_{13}$  at $1\sigma$ (top), $\frac{1}{2}\sigma$ (middle) 
 and very close to the central values at $\frac{1}{4}\sigma$ (bottom). All values of $\delta$ (except for those excluded by T2K~\cite{Abe:2013hdq}) are consistent and possible. Note that as $\sin^{2}\delta$ approaches zero the range for $\theta^{\ell}_{13}$ becomes $0.06 \leq \theta^{\ell}_{13} \leq 0.14$ at $1\sigma$ for NH ($0.06 \leq \theta^{\ell}_{13} \leq 0.12$ at $1\sigma$ for IH) and singles out the values $\theta^{\ell}_{13} \sim 0.08$ and $\theta^{\ell}_{13} \sim 0.12$ at the central values for NH ($\theta^{\ell}_{13} \sim 0.08$ and $\theta^{\ell}_{13} \sim 0.11$ for IH).}
  \label{fig4}
\end{figure}

\begin{figure}[ht]
%\begin{center}
  \includegraphics[width=8cm]{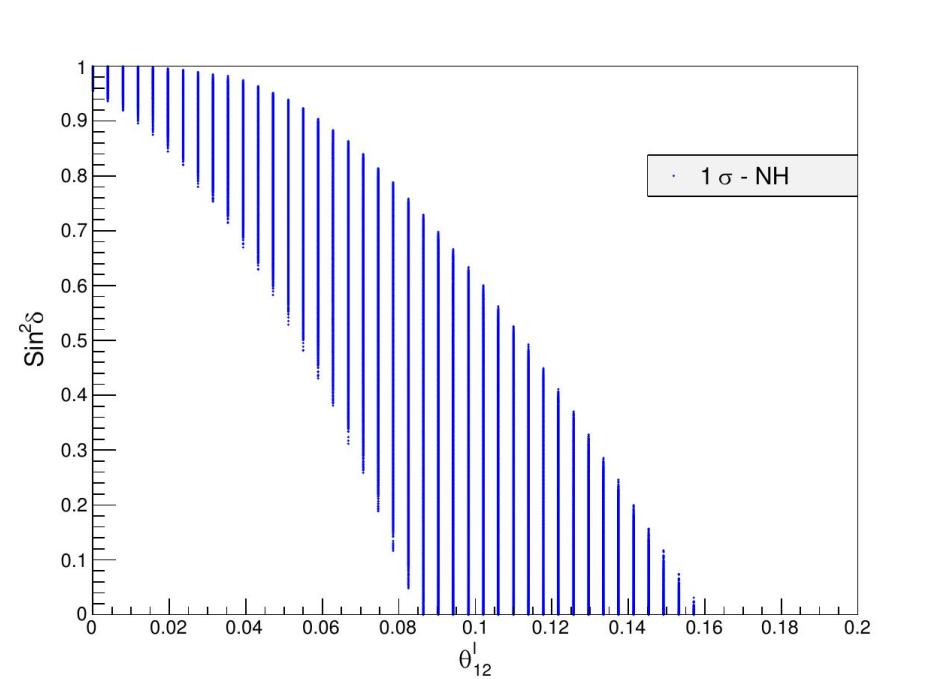}
 \includegraphics[width=8cm]{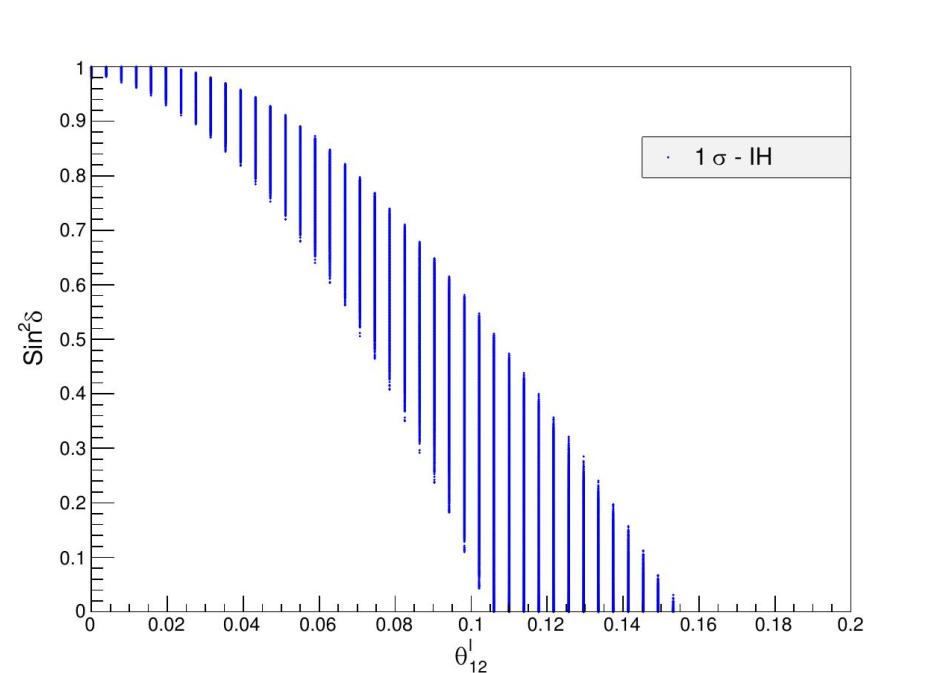} \\
 \includegraphics[width=8cm]{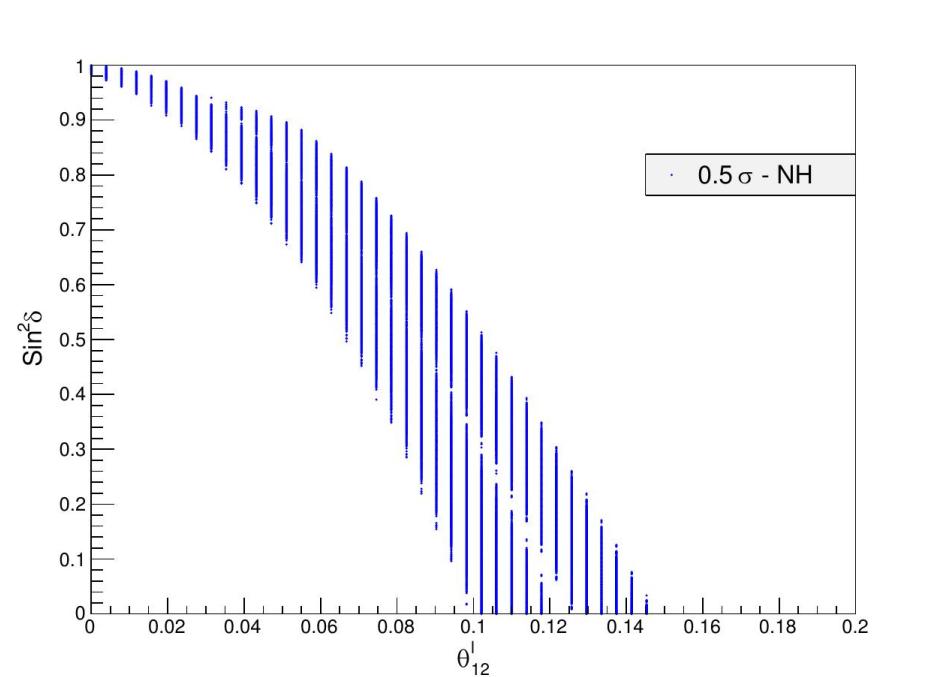}
 \includegraphics[width=8cm]{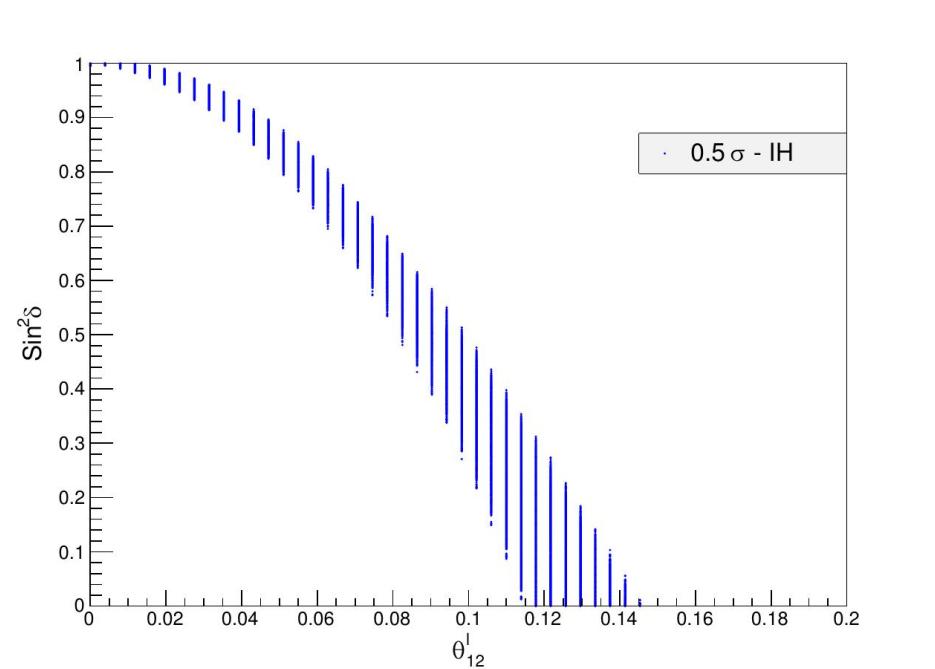} \\
 \includegraphics[width=8cm]{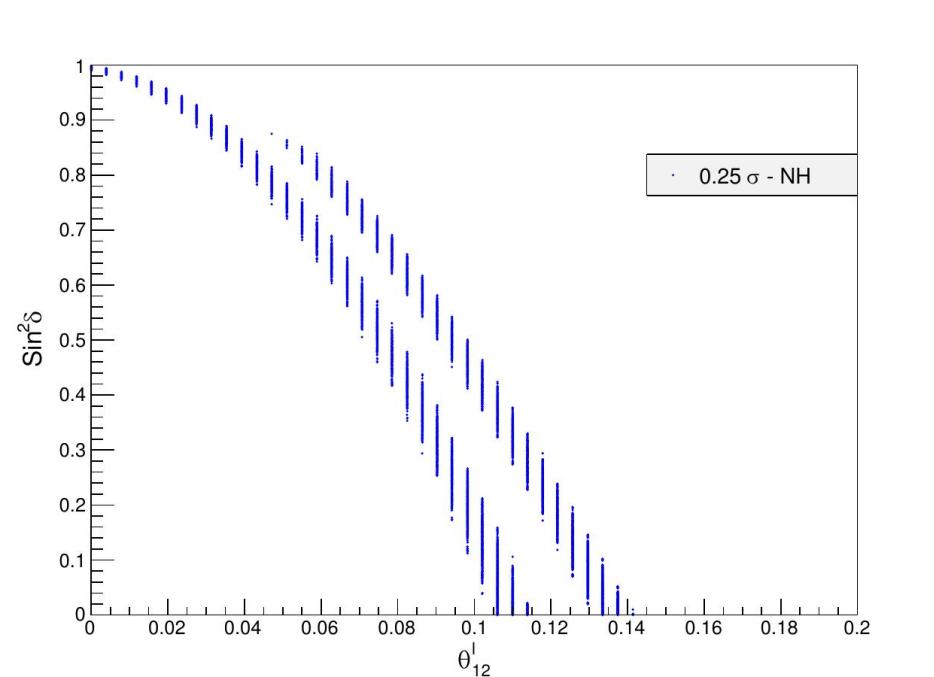}
 \includegraphics[width=8cm]{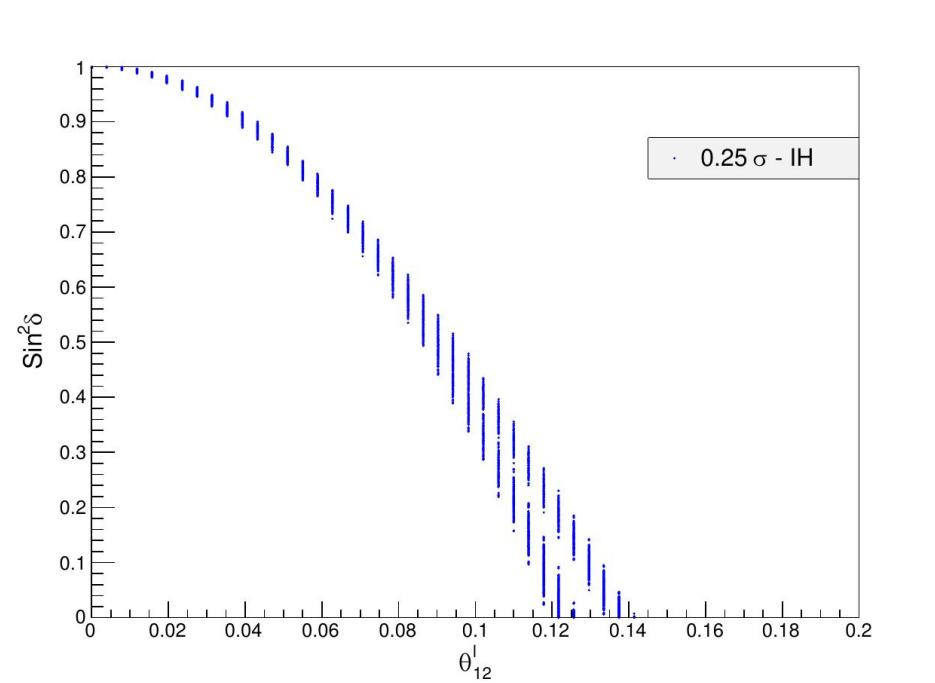}
 %\end{center}
 \caption{Variation of $\sin^{2}\delta$ with respect to $\theta^{\ell}_{12}$  at $1\sigma$ (top), $\frac{1}{2}\sigma$ (middle) 
 and very close to the central values at $\frac{1}{4}\sigma$ (bottom). All values of $\delta$ (except for those excluded by T2K~\cite{Abe:2013hdq}) are consistent and possible. Note that as $\sin^{2}\delta$ approaches zero the range for $\theta^{\ell}_{12}$ becomes $0.085 \leq \theta^{\ell}_{13} \leq 0.16$ at $1\sigma$ for NH ($0.105 \leq \theta^{\ell}_{13} \leq 0.155$ at $1\sigma$ for IH) and singles out the values $\theta^{\ell}_{13} \sim 0.105$ and $\theta^{\ell}_{13} \sim 0.14$ at the central values for NH ($\theta^{\ell}_{13} \sim 0.12$ and $\theta^{\ell}_{13} \sim 0.14$ for IH).}
  \label{fig5}
\end{figure}

Figures~\ref{fig4} and~\ref{fig5} show the relation between $\sin^2\delta$ and the charged lepton mixing angles $\theta^{\ell}_{13}$ and $\theta^{\ell}_{12}$. As expected, they show an {\it inverted} relation, i.e. small values of one correspond to large values of the other. The dependence on $\theta^{\ell}_{23}$ does not give any further information. Note that since we are computing $\sin^2\delta$ the excluded ranges for $\delta$ recently obtained by the T2K collaboration do not show up in these figures: given a value for $\sin^{2}\delta$ in the plots, there are four possible values of $\delta$ for NH and two for IH (the results obtained by the T2K collaboration for IH exclude $\sin\delta \geq 0$) consistent with that value and at least one of them is always outside the exclusion regions.

According to the behavior shown in Figures~\ref{fig4} and \ref{fig5}  the only way we can make a 
prediction about $\delta$ under our assumptions is by fixing $\theta^{\ell}_{13}$ or $\theta^{\ell}_{12}$. For instance, by fixing
$\theta^{\ell}_{12}=0$, we can find the results obtained in~\cite{pet} for the TBM case where $\sin^{2}\delta \sim 1$, i.e. setting $\theta^{\ell}_{12}=0$, 
means fixing $\delta^{\ell}$ around $\pi/2$ or $3\pi/2$ (see Figure~\ref{fig2}) and $\sin^2\delta \sim 1$ (Figure~\ref{fig5}). Thus, $\delta$ lies around $\pi/2$ or $3\pi/2$ (for both NH and IH). Taking the T2K recent bound~\cite{Abe:2013hdq} implies in this case that the only viable possibility is $\delta \sim 3\pi/2$ for both NH and IH and hence we obtain maximal CP-violation for leptons. This case corresponds to the $U_{\ell}$ matrices in Eqs.~\eqref{12min230} and~\eqref{12min2301}. For the case where $\theta_{13}$ gets its minimum value we obtain that $\sin\delta\sim 0$ and thus $\delta \sim 0,\pi$. Both solutions are allowed only for NH (this case corresponds to a $U_{\ell}$ of the form in Eq.~\eqref{13min}).

Another potentially interesting case from the model building point of view, is when $\theta^{\ell}_{12} \sim \theta^{\ell}_{13} \sim  \theta^{\ell}_{23}$. Taking as an example the central values for the IH case, we see that this occurs at around  $\theta^{\ell}_{12} \sim \theta^{\ell}_{13}\sim 0.13$ and $\theta^{\ell}_{23} \sim 0.15$. When this happens, the allowed values for $\delta^{\ell}$ are $\delta^{\ell} \sim 2.0$ and $\delta^{\ell} \sim 4.2$. Figures~\ref{fig4} and \ref{fig5}, together with the T2K exclusion ($\sin\delta < 0$) show that $\delta \in (-0.14\pi, -0.11\pi) \cup  (-0.86\pi, -0.89\pi) $. Such a possibility would require a charged lepton mixing matrix of the general form
\begin{eqnarray}
U_{\ell} \sim 
\left(
	\begin{array}{ccc}
	1 - \epsilon^2 & \epsilon & \epsilon e^{-i\delta_{\ell}} \\
	- \epsilon(1+ \epsilon e^{i\delta_{\ell}}) &  1 - \epsilon^{2}(1+ \epsilon e^{i\delta_{\ell}}) & \epsilon \\
	 \epsilon(\epsilon - e^{i\delta_{\ell}}) & -\epsilon(1 + \epsilon e^{i\delta_{\ell}}) & 1 - \epsilon^2 
	\end{array}
\right) \ ,
\end{eqnarray}
where $\epsilon ~ O(10^{-1})$. A model for lepton masses and mixings that automatically leads to this $U_{\ell}$ and the TBM form for $U_{\nu}$, will lead to a value for $\delta \in (-0.11\pi, -0.14\pi)\cup  (-0.86\pi, -0.89\pi) $. A final comment before concluding is that models leading to a sizeable values for $\theta^{\ell}_{23}$ might lead to interesting lepton flavor violating phenomenology~\cite{fv}.

\section{Conclusion}
We present a model independent analysis of the lepton sector CP violating Dirac phase, $\delta$, under the following general setup: the neutrino mixing matrix is exactly TBM and the non-zero value of $\theta_{13}$ in the PMNS breaking is obtained through the charged lepton mixing matrix $U_{\ell}$. This matrix is then parametrized in terms of three mixing angles $(\theta^{\ell}_{12}, \ \theta^{\ell}_{13}, \ \theta^{\ell}_{23})$  and one phase $\delta^{\ell}$. Using the most current fits and data we find that, except for the regions excluded at $90\%$ C.L. by T2K~\cite{Abe:2013hdq}, any value of $\delta$ is still consistent and that in order to obtain any {\it prediction} an additional assumption must be made. The allowed ranges for the $\theta^{\ell}_{ij}$ angles have been determined and an interesting feature is that $\theta^{\ell}_{23}$ can practically take values only in two narrow ranges around $0$ and $\pi/2$. The allowed values for the other two angles are more constrained and must be small. In particular we find that $\theta^{\ell}_{13}$, albeit small, cannot be zero. We singled out some special cases typically discussed in the literature. In addition we explored the value of $\delta$ for the interesting case where the charged lepton mixing angles are all similar in size and found that, if that is the case, $\delta \in (-0.11\pi, -0.14\pi)\cup  (-0.86\pi, -0.89\pi) $ for IH. A specific model leading to such a mixing matrix is an interesting possibility. Another potentially interesting case corresponds to models where the mixing in the $2-3$ sector is large and might lead to lepton flavor violating contributions. A study of these model dependent issues is currently being pursued.

\acknowledgments
J.A.A. acknowledges support  (DFG/GRK1581) through a fellowship from the GRK Symmetry Breaking. A.A. was supported in part by CONACYT and PROMEP. J.V. thanks the Programa  DELFIN for support during the Verano de la Investigaci\'on Cient\'ifica y Tecnol\'ogica del Pac\'ifico.

\bibliographystyle{ieeetr}

\end{document}